\documentclass[journal ]{new-aiaa}
\usepackage[utf8]{inputenc}
\usepackage{textcomp}

\usepackage{algorithm}
\usepackage{algorithmic}
\usepackage{makeidx}
\usepackage{tabularx}
\usepackage{enumerate}
\usepackage{bm}
\usepackage{wrapfig}
\usepackage{hhline}
\usepackage{url}
\usepackage{pifont}
\usepackage{amsmath}
\usepackage{mathtools}
\usepackage{fancyhdr}
\usepackage{here}
\usepackage{cases}
\usepackage{multirow}
\usepackage{ascmac}
\usepackage{booktabs}
\usepackage{color}
\usepackage{caption}
\usepackage{subcaption}
\usepackage{setspace}
\usepackage{graphicx}
\usepackage{xcolor}
\DeclareRobustCommand{\erase}{\bgroup\markoverwith{\textcolor{red}{\rule[.5ex]{2pt}{0.4pt}}}\ULon}

\usepackage[version=4]{mhchem}
\usepackage{siunitx}
\usepackage{longtable,tabularx}

\title{Attitude Control of Spacecraft for Autonomous Attenuation of Unknown Periodic Disturbance Torque}

\author{Yuta Hayashi \footnote{Graduate student, Department of Aeronautics and Astronautics, 744, Motooka, Nishi-ku, Fukuoka; hayashi.yuta.963@s.kyushu-u.ac.jp}, Mai Bando\footnote{Professor, Department of Aeronautics and Astronautics, 744, Motooka, Nishi-ku, Fukuoka; mbando@aero.kyushu-u.ac.jp, Senior Member AIAA.} and Shinji Hokamoto\footnote{Professor, Department of Aeronautics and Astronautics, 744, Motooka, Nishi-ku, Fukuoka; hokamoto@aero.kyushu-u.ac.jp. Member AIAA.}}
\affil{Kyushu University, Fukuoka 819-0395, Japan}
\begin{document}

\maketitle

\begin{abstract}
Recently, deep space exploration, especially focusing on halo orbits, the periodic orbits of the Moon, has been widely studied. The spacecraft in halo orbits performs periodic orbital motion, which affects the attitude motion by periodic disturbances. The conventional attitude control method, PD control, is widely used, but its application to periodic disturbance attenuation is inefficient. To address these challenges, this study proposes a predictive Repetitive Control (RC) approach that addresses periodic disturbances, particularly GG torque, by exploiting the periodic nature of the system dynamics. The proposed method is also applied to the case of using a Reaction Wheel (RW) as an attitude control actuator. Despite the inherent challenges posed by RW limitations, including saturation torque and transmission delay, our predictive RC approach effectively mitigates these effects. Numerical simulations demonstrate the robust performance of the proposed method in maintaining attitude control for spacecraft traversing halo orbits near the Earth-Moon $L_2$ point, validating its potential for future deep space exploration missions.
\end{abstract}

\section{Introduction}
With recent advancements in deep space exploration, halo and Lyapunov orbits around the Earth–Moon Lagrangian point (L$_2$) in the Circular Restricted Three-Body Problem (CRTBP) have garnered significant attention from both scientific and engineering communities~\cite{kane1971attitude,Robinson1974,abad1989attitude,brucker2007analysis,wong2008attitude}. These orbits offer stable platforms for long-duration missions, enabling wide-area observation and enhancing deep-space communication infrastructure. However, their inherent instability and sensitivity to small numerical errors necessitate high-precision analysis and control to ensure long-term mission feasibility \cite{guzzetti2017natural,knutson2015attitude,colagrossi2021guidance}.
Traditionally, spacecraft attitude control has been designed independently of orbit control under the assumption that disturbance torques are significantly smaller than orbital perturbations. However, in deep space missions, precise attitude control is critical for maintaining antenna pointing accuracy and stabilizing observation instruments—both essential for mission success. When attitude dynamics directly influence orbital motion, the coupling between attitude and orbit dynamics introduces significant complexity into the spacecraft behavior. As future missions become increasingly sophisticated, there is a growing demand for an integrated “orbit-attitude coupled model”~\cite{kikuchi2019stability,kikuchi2017orbit} that simultaneously captures both dynamics.

The importance of such a coupled model has been demonstrated in advanced exploration missions, such as sample return missions. For example, JAXA's Hayabusa2 successfully conducted precise orbit and attitude stability analysis near an asteroid by employing an orbit–attitude coupled model that accounted for gravity-gradient (GG) torque and solar radiation pressure. In this mission, the analysis is performed within the two-body problem framework, allowing analytical treatment of the coupled dynamics. In contrast, in multi-body environments such as the Earth–Moon CRTBP, the nonlinearity and lack of closed-form solutions significantly increase the complexity of coupled dynamics analysis. This renders the simultaneous design and control of orbit–attitude interactions particularly challenging. As a result, developing new methods for quantitative stability evaluation and numerical analysis remains an open and critical research issue.

This paper focuses on the attitude motion of a spacecraft orbiting along a halo orbit near the Earth-Moon L$_2$ point. As an initial step toward fully integrated analysis, a “one-way coupled model” is employed, in which the influence of orbital motion on attitude dynamics is considered, while feedback from attitude to orbit is neglected. This simplification enables the avoidance of the significant computational burden associated with high-dimensional, fully coupled systems, while still capturing the essential effects of orbit-induced disturbances on attitude motion.

Conventional attitude control methods for spacecraft include PD control \cite{schlanbusch2012pd+}, adaptive control\cite{Luo2005,Ahmed1998}, and sliding mode control \cite{An2011}. In recent years, advanced control methods such as learning control, optimal control \cite{sharma2004optimal}, and model predictive control \cite{hartley2015tutorial,Christopher2016} have been explored to enhance spacecraft autonomy and robustness. However, these control methods are not tailored for periodic disturbances, limiting their effectiveness in deep space missions where gravitational perturbations exhibit strong periodicity.

For a spacecraft moving along a halo orbit, GG torque from Earth and Moon constitutes major disturbances that destabilize attitude motion. Accurately modeling these torques is challenging due to higher-order gravitational effects. Therefore, it is essential to consider the uncertainty of such disturbances, and control approaches that rely solely on input-output signals rather than full dynamic models are particularly advantageous in these environments.

To address these challenges, this study proposes an anticipatory Repetitive Control (RC) approach to mitigate periodic disturbance torques, particularly GG torque, by leveraging the inherent periodicity of the system dynamics. While Iterative Learning Control (ILC) \cite{arimoto1984bettering} has been widely studied for both periodic and non-periodic systems, RC \cite{19861111,aida1984design,hara1988repetitive} offers distinct advantages in periodic environments by directly learning the disturbance pattern over successive cycles\cite{198636,hara1988repetitive,1986830}. This property enables RC to achieve higher accuracy in compensating for periodic disturbances. A key limitation of ILC is that it requires resetting control inputs to initial conditions after each iteration, making it impractical for real-time spacecraft operations. In contrast, RC continuously updates the control input in an online manner, making it well-suited for long-duration space missions.

Building upon the anticipatory framework developed by Wu et al. \cite{wu2015high}, the proposed anticipatory RC method utilizes future error predictions instead of relying solely on past data. This anticipatory mechanism enhances disturbance rejection performance and provides robust attitude control in the presence of modeling uncertainties.

Furthermore, this study considers the use of Reaction Wheels (RWs) as attitude control actuators. Despite inherent limitations such as torque saturation and transmission delays, the proposed anticipatory RC approach demonstrates strong robustness against these nonlinearities. Moreover, the attitude control is extended to scenarios involving the orbit transfer problem, where the reference trajectory evolves continuously over time. The proposed control strategy enables the spacecraft to maintain stable attitude behavior even under dynamically changing orbital conditions. Numerical simulations are conducted to verify the effectiveness of the proposed method in maintaining precise attitude control along halo orbits near the Earth–Moon L$_2$ point. The results highlight the method’s potential for enhancing the reliability of future deep space exploration missions.

The remainder of this paper is organized as follows. Section 2 formulates the orbit-attitude coupled dynamical model of a spacecraft moving in the halo orbits. Section~3 presents the proposed attitude control law based on RC, designed to attenuate unknown periodic disturbance torques under uncertainty. Section~4 validates the control performance through numerical simulations, demonstrating the effectiveness of the RC scheme in the presence of RW dynamics and dynamically changing reference orbits. Finally, Section~5 concludes the paper and discusses future research directions.
\section{Dynamical model}
\begin{figure}[htbp]
 \begin{center}
    \includegraphics[scale=0.5]{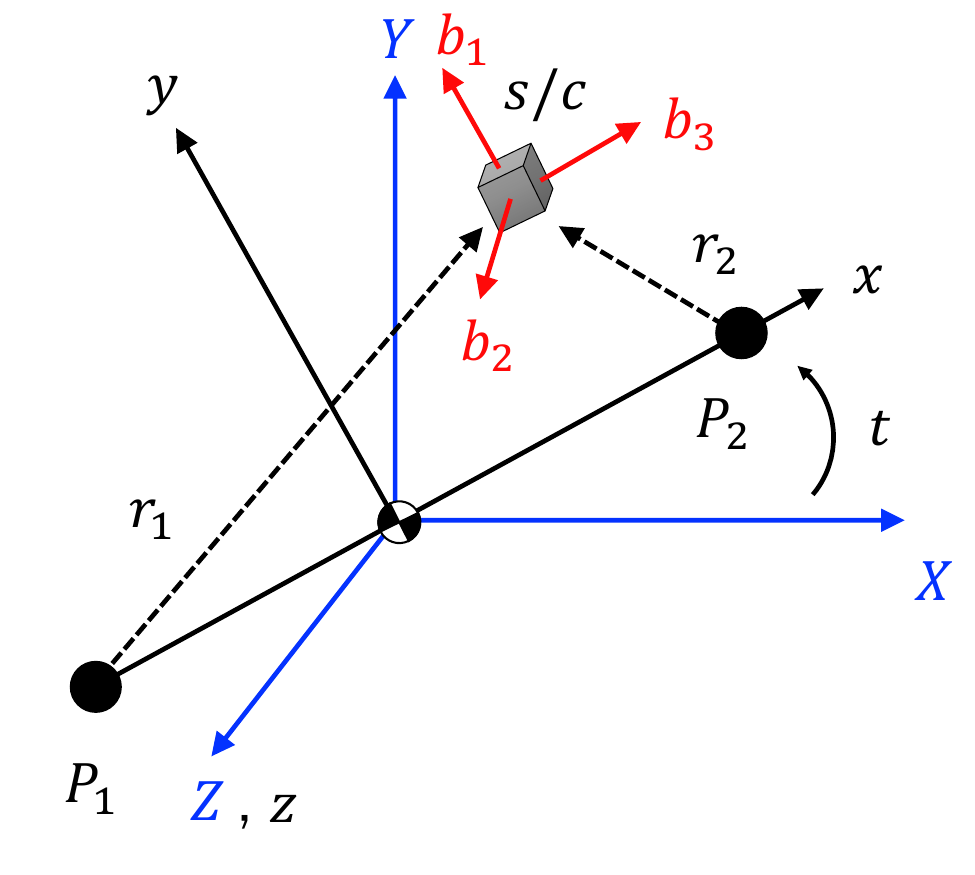}
	\caption{Geometry of the CRTBP}%
	\label{fig:coordinate_system}
 \end{center}
\end{figure}

Consider the motion of a spacecraft moving under the gravitational attraction from $P_1$ (Earth) and $P_2$ (Moon), the masses of the Earth and the Moon are denoted as $m_1$ and $m_2$, respectively. Let $\bm{r}_1$ and $\bm{r}_2$ be the position vectors of the spacecraft relative to the Earth and the Moon, respectively.
Let ($\bm{X}, \bm{Y}, \bm{Z}$) be the inertial frame whose origin is the barycenter of the system, and ($\bm{x}, \bm{y}, \bm{z}$) be the rotating frame with the barycenter of the system as the origin and the $\bm{x}$-axis is defined by the direction vector from the primary to the secondary body and (${\bm{b}_1}, {\bm{b}_2}, {\bm{b}_3}$) be the body-fixed frame whose origin is at the center of mass of a spacecraft (see Fig.~\ref{fig:coordinate_system}). 
Assuming that a spacecraft's mass is sufficiently small compared to $P_1$ and $P_2$ and that the secondary moves in a circular motion around the primary, the problem is modeled by the CRTBP. 

\subsection{Equations of motion}
\subsubsection{Orbital Dynamics}
The dimensionless Earth-Moon mass parameter ${\mu}$ is expressed as ${\mu} = \frac{m_2}{m_1 + m_2}$, where $m_1$ and $m_2$ are the masses of the Earth and the Moon, respectively.
The equations of motion of the CRTBP in a rotating frame whose origin is at the barycenter of the system, with coordinates $(\bm{x},\bm{y},\bm{z})$, are given by
\begin{equation}
    \begin{split}        
	\ddot{x}&= x + 2\dot{y}-\frac{(1-\mu)(x+\mu)}{r_1^3}-\frac{\mu(x-1+\mu)}{r_2^3}+a_{p,x}\\
	\ddot{y}&=y -2\dot{x}-\frac{(1-\mu)y}{r_1^3}-\frac{\mu y}{r_2^3}+a_{p,y}\\
	\ddot{z}&= -\frac{(1-\mu)z}{r_1^3}-\frac{\mu z}{r_2^3}  +a_{p,z}
    \end{split}
    \label{eq:crtbp}
\end{equation}
where $a_{p,j}$ (with $j=x,y,z$) represents acceleration perturbations due to second-order gravitational terms and other perturbations such as atmospheric drag and tidal forces, which arise when considering the finite extent of the spacecraft. However, in this paper, we neglect these contributions as they are considered negligible.
\par
\begin{figure}[htbp]
  \centering
    \includegraphics[scale=0.4]{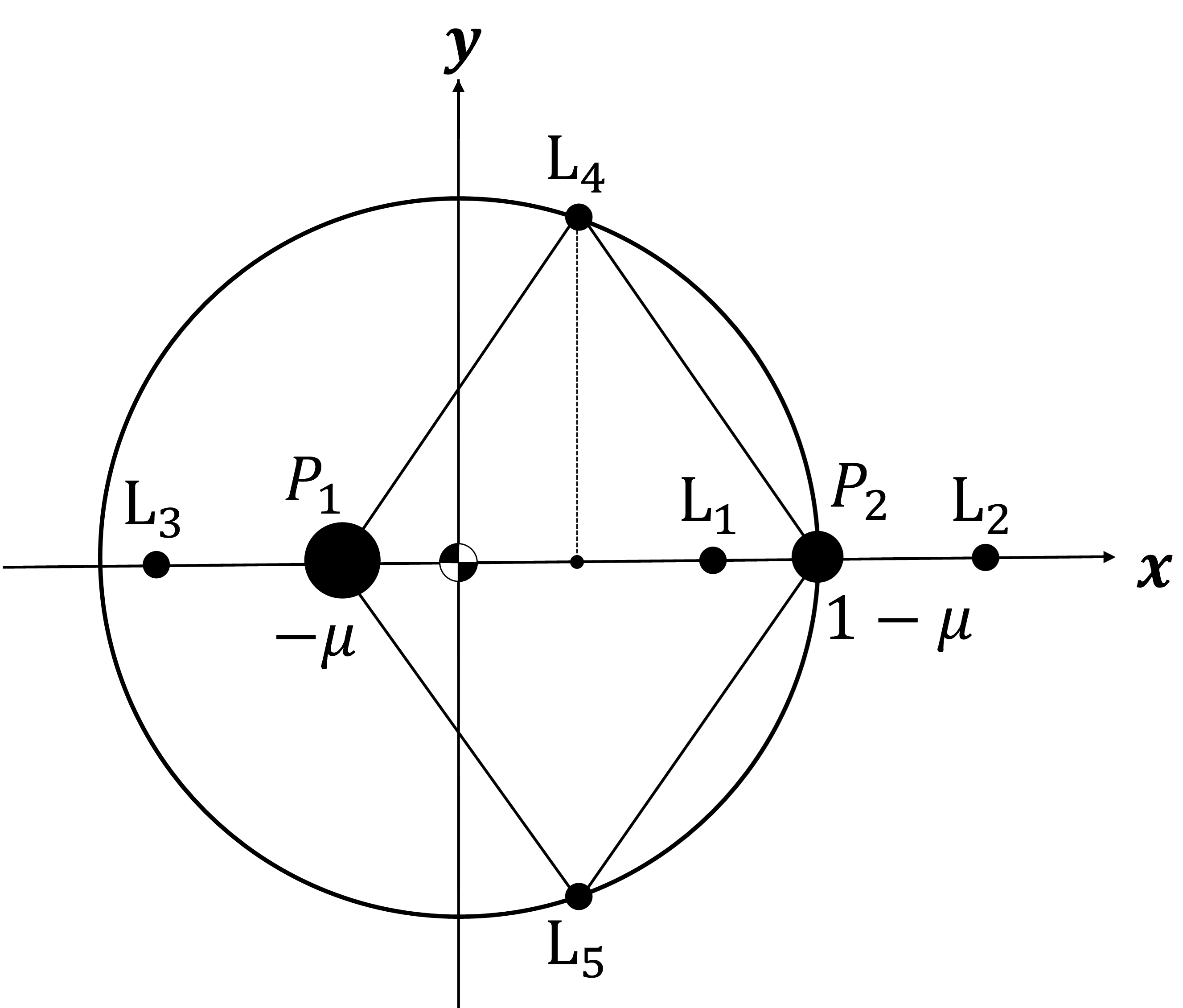}
	\caption{Lagrange point}%
	\label{fig:Lagrangian_Point}
\end{figure}
\begin{figure}[htbp]
  \centering
    \includegraphics[scale=0.55]{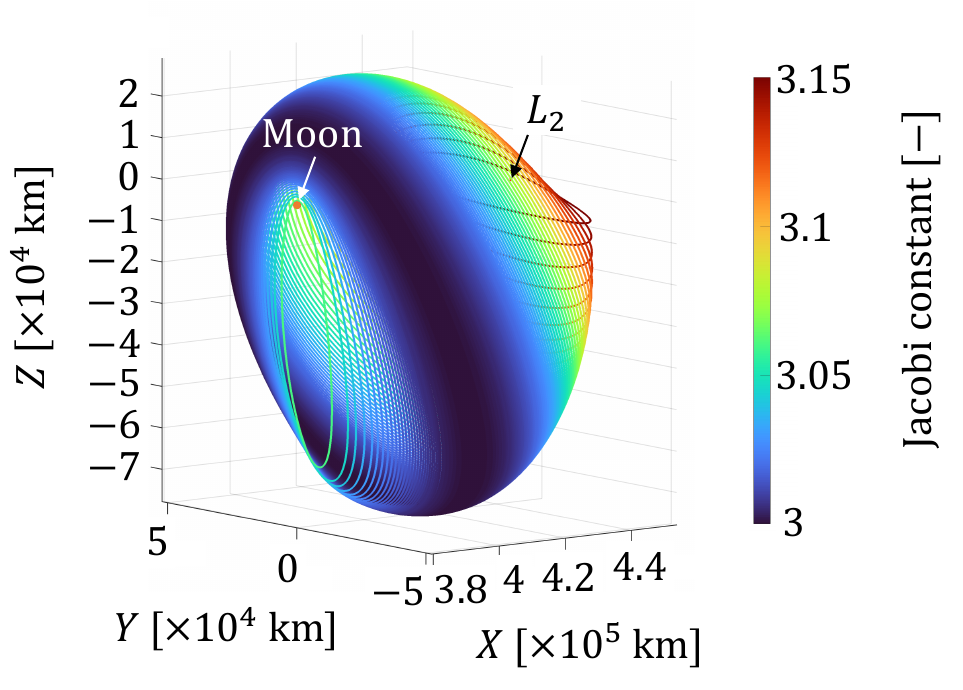}
	\caption{L$_2$ halo orbits}%
	\label{fig:coordinate_halo}
\end{figure}
\begin{figure}[htbp]
  \centering
    \includegraphics[scale=0.5]{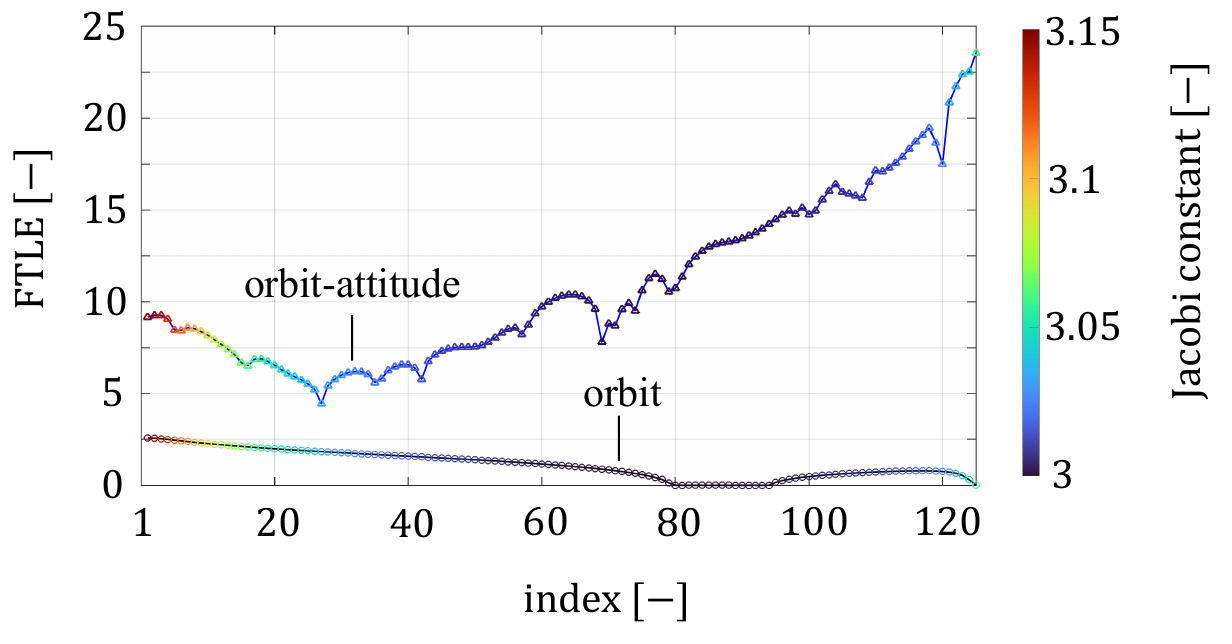}
	\caption{FTLE for each halo orbit from near the L2 point to the Moon~~($J'=\mathrm{diag}(135,175,125) [\mathrm{kgm^2}])$}%
	\label{fig:FTLE}
\end{figure}

In addition, there exists a conserved quantity called the Jacobi constant, which is described as follows: 
\begin{align}
C&={x^2}+{y^2}+\cfrac{2(1-\mu)}{r_1}+\cfrac{2\mu}{r_2}-({\dot x}^2+{\dot y}^2+{\dot z}^2) \label{Jacobi}
\end{align}
The equilibrium point of Eq.~\eqref{eq:crtbp} is called the Lagrange point as shown in Fig.~\ref{fig:Lagrangian_Point}, where gravitational and centrifugal forces balance.
Figure~\ref{fig:coordinate_halo} shows the halo orbits near the Earth-Moon L$_2$ point, which are used as reference orbits in this paper. The color bar indicates the Jacobi constant for each orbit, and the orbits are colored accordingly. Furthermore, it is known that there exists a family of planar periodic orbits called Lyapunov orbits and a family of spatial periodic orbits called halo orbits around the collinear libration points in Fig.~\ref{fig:coordinate_halo}. 
Figure~\ref{fig:FTLE} shows the Finite-Time Lyapunov Exponent (FTLE), which quantitatively represents the instability of a spacecraft moving along the halo orbits shown in Fig.~\ref{fig:coordinate_halo}, for both orbital motion and the orbit-attitude coupled model.
In the analysis of the coupled motion, the GG torque that primarily destabilizes the attitude motion depends on the inertia moment ratio of the spacecraft. Therefore, the instability of the coupled model is evaluated for a spacecraft with a representative moment of inertia of the spacecraft $J'$. The initial attitude of the spacecraft is assumed to be aligned with the inertial frame, and the attitude motion is assumed to be free.

As a result, it is confirmed that the instability of the coupled model considering attitude motion increases drastically compared to the orbital motion alone. This increase implies that the attitude motion is strongly affected by the disturbance torques from the two celestial bodies, which makes the overall dynamics more complex. Therefore, properly analyzing the coupling effect between orbit and attitude is essential.

\subsubsection{Attitude Dynamics}
The attitude motion of a spacecraft follows Euler's equation of rotational motion as follows:
\begin{equation}
       J\dot{\bm{\omega}}(t)+{\omega}(t)^{\times}J\bm{\omega}(t) = \bm{u}(t)+\bm{T}_{\mathrm{ext}}(t)
       \label{eq:Attitude Dynamics}
\end{equation}
where $J$ is the inertia moment of a spacecraft and $\bm{\omega}(t)=[\omega_1(t),\omega_2(t),\omega_3(t)]^T$ is the angular velocity vector. $\bm{u}(t)=[u_1,u_2,u_3]^T$ is the control input vector derived from the attitude control system of a spacecraft, which in this paper is the control torque generated by attitude actuators such as RWs. $\bm{T}_{\mathrm{ext}}(t)$ is the disturbance torque and the cross-product ($^\times$) is defined by the following skew-symmetric matrix.
\begin{equation}
    {a}^{\times} =
    \begin{bmatrix}
        0&-a_3&a_2\\a_3&0&-a_1\\-a_2&a_1&0
    \end{bmatrix}  
\end{equation}
For the disturbance torque $\bm{T}_{\mathrm{ext}}(t)$, we consider only the GG torque, a periodic disturbance torque from the Earth and Moon, expressed as:
\begin{equation}
    \bm{T}_{\mathrm{ext}}(t)=
\frac{3(1-\mu)}{r_1^5}(-{g}(t)^\times J\bm{g}(t))+\frac{3\mu}{r_2^5}(-{h}(t)^\times J\bm{h}(t))
    \label{eq:disturbance}
\end{equation}
where $\bm{g}(t)$ and $\bm{h}(t)$ are defined using the Direction Cosine Matrix (DCM) and the spacecraft’s position vector as follows.
\begin{equation}
    \bm{g}(t)=
    \left[
    \begin{aligned}
        g_1(t)\\
        g_2(t)\\
        g_3(t)
   \end{aligned}
   \right]=
[A_{b/i}(t)][A_{i/r}(t)]\bm{r_1},
    \qquad
    \bm{h}(t)=
    \left[
    \begin{aligned}
        h_1(t)\\
        h_2(t)\\
        h_3(t)
    \end{aligned}
\right]=
[A_{b/i}(t)][A_{i/r}(t)]\bm{r_2}
\end{equation}
The DCM $[A_{i/r}(t)]$ converts vectors from the rotating frame to the inertial frame, while $[A_{b/i}(t)]$ converts vectors from the inertial frame to the body-fixed frame of the spacecraft. These matrices are given by: 
    \small
\begin{equation}
    \setlength{\arraycolsep}{3pt} 
    \renewcommand{\arraystretch}{0.8} 
    \begin{split}
    A&_{b/i}(t)\\ = &
        \begin{bmatrix}
            q_1(t)^2-q_2(t)^2-q_3(t)^2+q_4(t)^2&2(q_1(t)q_2(t)+q_3(t)q_4(t))&2(q_1(t)q_3(t)-q_2(t)q_4(t))\\
            2(q_1(t)q_2(t)-q_3(t)q_4(t))&-q_1(t)^2+q_2(t)^2-q_3(t)^2+q_4(t)^2&2(q_2(t)q_3(t)+q_1(t)q_4(t))\\
            2(q_1(t)q_3(t)+q_2(t)q_4(t))&2(q_2(t)q_3(t)-q_1(t)q_4(t))&-q_1(t)^2-q_2(t)^2+q_3(t)^2+q_4(t)^2
        \end{bmatrix}
    \end{split}
    \label{eq:A_bi}
\end{equation}
\normalsize
\begin{equation}
    A_{i/r}(t)=
    \begin{bmatrix}
        \cos(t)&-\sin(t)&0\\
        \sin(t)&\cos(t)&0\\
        0&0&1
    \end{bmatrix}
\end{equation}
\subsubsection{Attitude Kinematics}
The kinematics of attitude motion can be described using a rotation axis $\bm{e} = [e_1, e_2, e_3]^T$ and a rotation angle $\theta$, or equivalently using quaternions.
The attitude quaternion is defined as:
\begin{equation}
        \begin{bmatrix}
            {\bm{q}(t)}\\
            {q_4(t)}
        \end{bmatrix}=
        \begin{bmatrix}
        \bm{e}\sin\frac{\theta}{2}\\
        \cos\frac{\theta}{2}
        \end{bmatrix}
    \label{eq:kinematics}
\end{equation}
where $\bm{q}(t)=[q_1(t),q_2(t),q_3(t)]^T$ and the following constraints must be satisfied. 
\begin{equation}
   q_1(t)^2+q_2(t)^2+q_3(t)^2+q_4(t)^2=1
    \label{eq:const}
\end{equation}
The quaternion kinematic equations body-fixed frame are given by
\begin{equation}
    \begin{bmatrix}
        \dot{\bm{q}}(t)\\
        \dot{q_4}(t)
    \end{bmatrix}=\frac{1}{2}
    \begin{bmatrix}
        q_4(t){I}_3+{q}(t)^{\times}\\
        -\bm{q}(t)^T
    \end{bmatrix}[\bm{\omega}(t)]
\end{equation}
where $I_3$ is a unit matrix of $3\times 3$ unit matrix.
\subsubsection{Attitude Error Dynamics}
The attitude error dynamics and kinematics are described as follows:
\begin{equation}
    J\delta\dot{\bm{\omega}}(t)=-{\omega}(t)^{\times}J\bm{\omega}(t)+J\{\delta{\omega}(t)^{\times}R[\delta\bar{q}(t)]\bm{\omega}_d(t)-R[\delta\bar{q}(t)]\bm{\dot{\omega}}_d(t)\}+\bm{u}(t)+\bm{T}_{\mathrm{ext}}(t)
    \label{eq:error dynamics}      
\end{equation}
\begin{equation}
    \begin{split}
        \begin{bmatrix}
            \delta\dot{\bm{q}}(t)\\
            \delta\dot{q_4}(t)
        \end{bmatrix}&=\frac{1}{2}
        \begin{bmatrix}
        \delta q_4(t)I_3+{q}(t)^{\times}\\
        -\delta\bm{q}(t)^T
        \end{bmatrix}[\delta\bm{\omega}(t)]
    \end{split}
    \label{eq:error kinematics}
\end{equation}
where $\delta\bm{\omega}(t)$ denotes the error angular velocity derived by the following equation, and $R$ is the rotation matrix. $\delta\bm{\omega}(t)$ is the error angular velocity and $\delta\bar{q}(t)=[\delta\bm{q}(t)^T,\delta q_4(t)]^T$ is the error quaternion, defined with respect to their target values $\bm{\omega}_d(t)$ and $[\bm{q}_d(t), q_{4,d}(t)]^T$ as follows:
\begin{equation}
\delta\bm{\omega}(t)=\bm{\omega}(t)-R[\delta\bar{q}(t)]\bm{\omega}_d(t)
\end{equation}
\begin{equation}
    \begin{split}
        \begin{bmatrix}
            \delta{\bm{q}}(t)\\
            \delta{q_4}(t)
        \end{bmatrix}=       
        \begin{bmatrix}
            \delta{q_1}(t)\\
            \delta{q_2}(t)\\
            \delta{q_3}(t)\\
            \delta{q_4}(t)
        \end{bmatrix}&=
        \begin{bmatrix}
        q_{4,d} & q_{3,d} & -q_{2,d} & -q_{1,d}\\
        -q_{3,d} & q_{4,d} & q_{1,d} & -q_{2,d}\\
        q_{2,d} & -q_{1,d} & q_{4,d} & -q_{3,d}\\
        q_{1,d} & q_{2,d} & -q_{3,d} & q_{4,d}
        \end{bmatrix}
        \begin{bmatrix}
            {\bm{q}}(t)\\
            {q_4}(t)
        \end{bmatrix}
    \end{split}
    \label{eq:error quaternion}
\end{equation}
\section{Controller design}
The attitude control law $\bm{u}(t)$ [Eq.~\eqref{eq:error dynamics}] is designed by combining a PD control, which is a classical feedback control with RC as follows:
\begin{equation}
    \bm{u}(t)=\bm{u}^{\mathrm{fb}}(t)+\bm{u}^{\mathrm{rc}}(t)
    \label{eq:input}
\end{equation}
The feedback input is effective for attenuating system errors and non-periodic disturbances, and the RC input is effective for attenuating periodic disturbances. Therefore, the control input [Eq.~\eqref{eq:input}] can provide a unified approach to various disturbances.

\subsection{Feedback control}
The PD control is used to guarantee closed-loop stability, and the control law is described as follows:
\begin{equation}
    \bm{u}^{\mathrm{fb}}(t)=-k_p^{\mathrm{fb}}J\delta \bm{q}(t)-k_d^{\mathrm{fb}}J\delta\bm{\omega}(t)+{\omega}(t)^{\times}J\bm{\omega}(t)+JR[\delta\bar{q}(t)]\bm{\dot{\omega}}_d(t)   
\label{eq:Feedback}
\end{equation}
where $k_p^{\mathrm{fb}}$ and $k_d^{\mathrm{fb}}$ are the PD gains. In the case of non-periodic disturbances, the errors of quaternion and angular velocity converge to zero under the PD control. Substituting [Eq.~\ref{eq:Feedback}] into [Eq.~ref{eq:error dynamics}] yields the following expressions for Eqs.~(\ref{eq:error kinematics}) and (\ref{eq:Feedback}):
\begin{equation}
        \begin{bmatrix}
            \delta\dot{\bm{q}}(t)\\\delta\dot{\bm{\omega}}(t)
        \end{bmatrix}=
        \begin{bmatrix}
        \frac{1}{2}[\delta q_{4}(t)I_3+\delta{q}(t)^{\times}]\delta\bm{\omega}(t)\\
        -k_p^{\mathrm{fb}}\delta \bm{q}(t)-k_d^{\mathrm{fb}}\bm{\omega}(t)
        \end{bmatrix}+
        \begin{bmatrix}
            0_{3\times3}\\J^{-1}
        \end{bmatrix}\bm{u}^{\mathrm{rc}}(t)+
        \begin{bmatrix}
            0_{3\times3}\\J^{-1}
        \end{bmatrix}\bm{T}_{\mathrm{ext}}(t)
    \label{eq:New dynamics and kinematics}
\end{equation}
\subsection{Repetitive control}
The RC adopts a feedforward-type control strategy that updates the control input by referencing error information obtained in previous trials. In this study, the RC input is updated for each trial of the spacecraft's orbital motion according to the following update law:
\begin{equation}
    \bm{u}^{\mathrm{rc}}(t)=\bm{u}^{\mathrm{rc}}(t-T)-k_p^{\mathrm{rc}}J\delta\bm{q}(t+\Delta-T)-k_d^{\mathrm{rc}}J\delta\bm{\omega}(t+\Delta-T)
    \label{eq:RC scheme}
\end{equation}
where $k_p^{\mathrm{rc}}$ and $k_d^{\mathrm{rc}}$ denote the proportional and derivative gains of the RC, respectively. The terms $\delta\bm{q}(t)$ and $\delta\bm{\omega}(t)$ represent the deviations of the quaternion and angular velocity from their target values.
In the first trial ($0 \le t \le T$), the RC input $\bm{u}^{\mathrm{rc}}(t)$ is initialized to zero.\\
This control law reduces the deviation from the target values in subsequent trials by correcting errors based on the information obtained in previous iterations.
In addition, this study introduces a novel anticipatory factor $\Delta$ into the classical RC framework to improve convergence accuracy.
Although the anticipatory ILC proposed in a previous study \cite{wu2015high} requires the complete reset of all state variables at the beginning of each trial, the proposed anticipatory RC can be applied without continuously resetting the variables.
Moreover, in the presence of periodic disturbances and model uncertainties, the boundedness of both control inputs and error states can be guaranteed if the following inequality is satisfied as follows:
\begin{equation}
    0<||1-\Delta\cdot k^{\mathrm{RC}}_d||<1
    \label{eq:theorem_rc}
\end{equation}
where $\Delta$ is chosen to be small enough.
The proof of the stability conditions \eqref{eq:theorem_rc} is similar to that of the Theorem of anticipatory ILC\cite{wang2009survey}. and can be derived through a unified mathematical formulation.

\section{Numerical Simulation}
In this section, the effectiveness of RC is evaluated through two numerical simulations for the orbit transfer problem and the case considering the dynamics of the actuator.
\subsection{Orbit transfer problem}
\begin{figure}[H]
  \centering
    \includegraphics[scale=0.45]{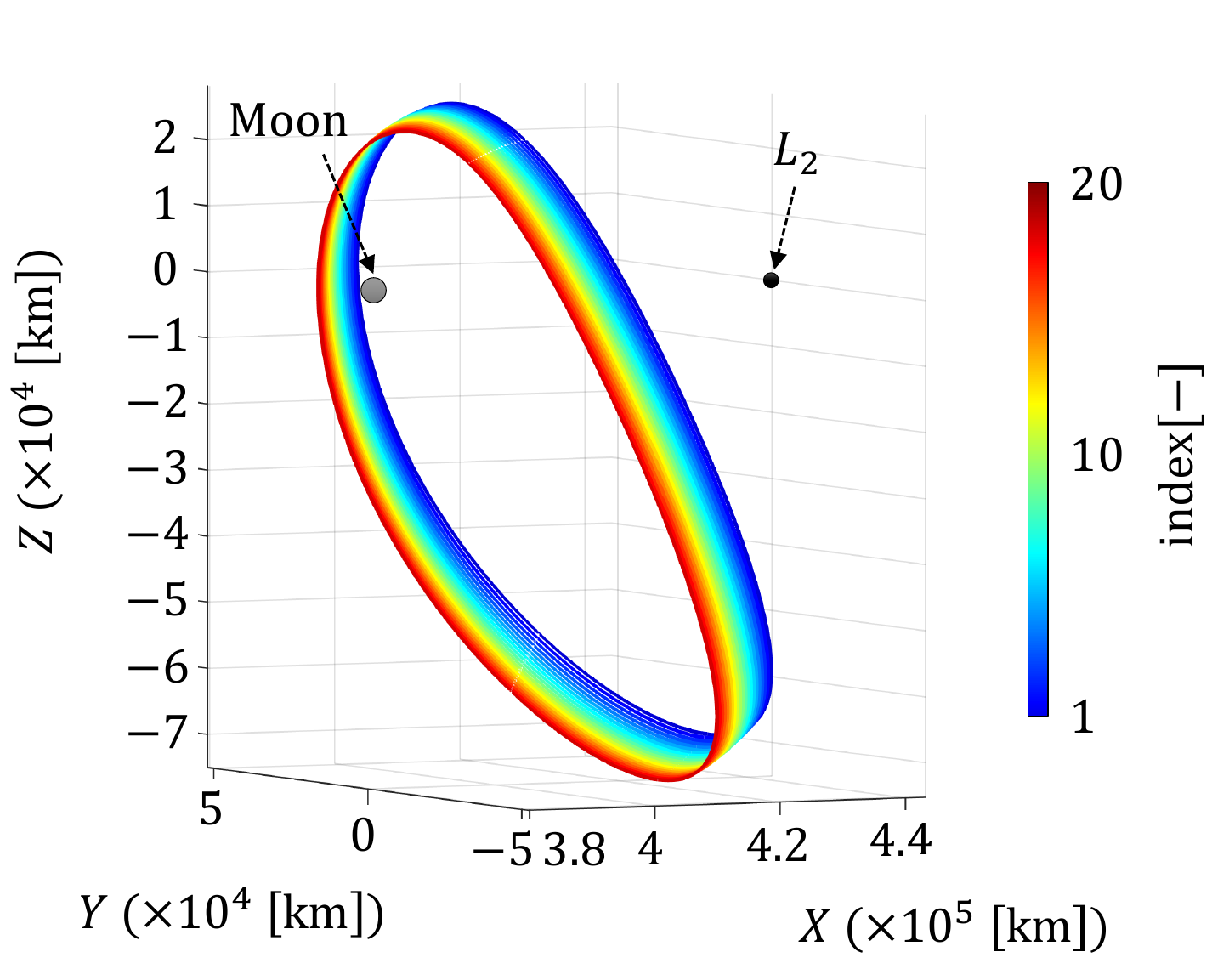}
	\caption{Change of reference orbit}%
	\label{fig:halofamily}
\end{figure}

In this section, attitude control of a spacecraft in the L$_2$  halo family in Fig.~\ref{fig:coordinate_halo} is considered. Table \ref{tb:parameter of spacecraft} shows the spacecraft parameters and state quantities used in the simulations. The information about the halo orbits referred to for the orbit transfer and the anticipatory parameter are also shown in Table \ref{tb:parameter of numerical analysis}. For the reference orbits, it is assumed that the spacecraft transfers a total of 20 discrete orbits from index 1 to 20 shown in Fig.~\ref{fig:halofamily}. For each index, the number of trials is changed. The number of trials is increased in the initial trials to sufficiently learn the periodicity and dynamic characteristics of the GG torque from the Earth and the Moon.
The control gains for PD control and RC are shown in Table \ref{tb:gain parameters}. The proportional gain is increased in order to improve the speed of convergence to the target value by ensuring a faster response to errors, and the differential gain is set larger to reduce system oscillations caused by the large proportional gain. As for the RC gains, the gains are set to higher values to improve the learning rate with fewer trials. These control gains are determined by trial and error.\par
\begin{figure}[H]
	\centering
		\includegraphics[width=0.85\hsize]{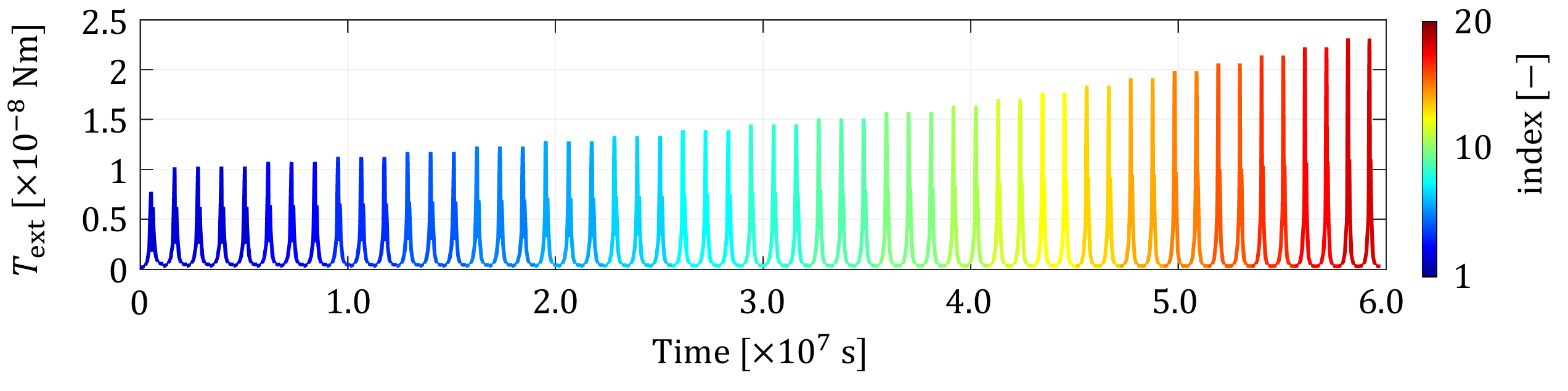}
		\caption{Time history of GG torque norm $\bm{T}_{\mathrm{ext}}$}
		\label{fig:GGT_transfbrc}
	\label{fig:trans_GGT}
\end{figure}
\begin{figure}[htbp]
    \begin{subfigure}{1\linewidth}
    	\centering
		\includegraphics[width=0.85\hsize]{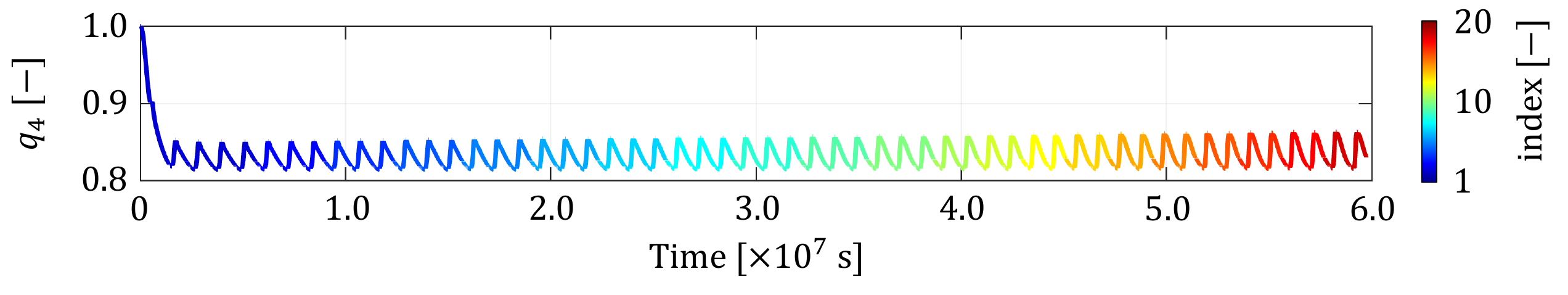}
		\caption{PD control}
		\label{fig:q4_transfb}
	\end{subfigure}
     \begin{subfigure}{1\linewidth}
     	\centering
		\includegraphics[width=0.85\hsize]{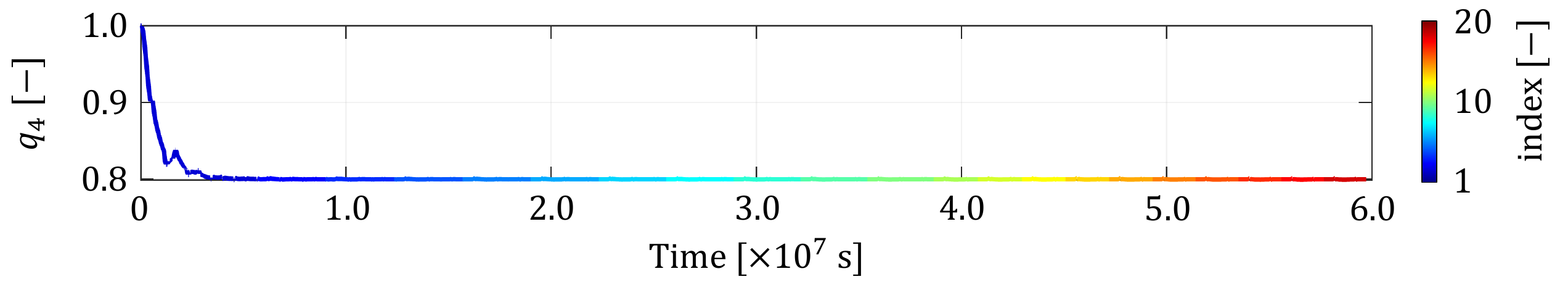}
		\caption{PD control + RC}
		\label{fig:q4_transfbrc}
	\end{subfigure}
	\caption{Time history of $q_4$.}
	\label{fig:trans_quaternion}
\end{figure}
\begin{figure}[htbp]
	\begin{subfigure}{1\linewidth}
	\centering
		\includegraphics[width=0.85\hsize]{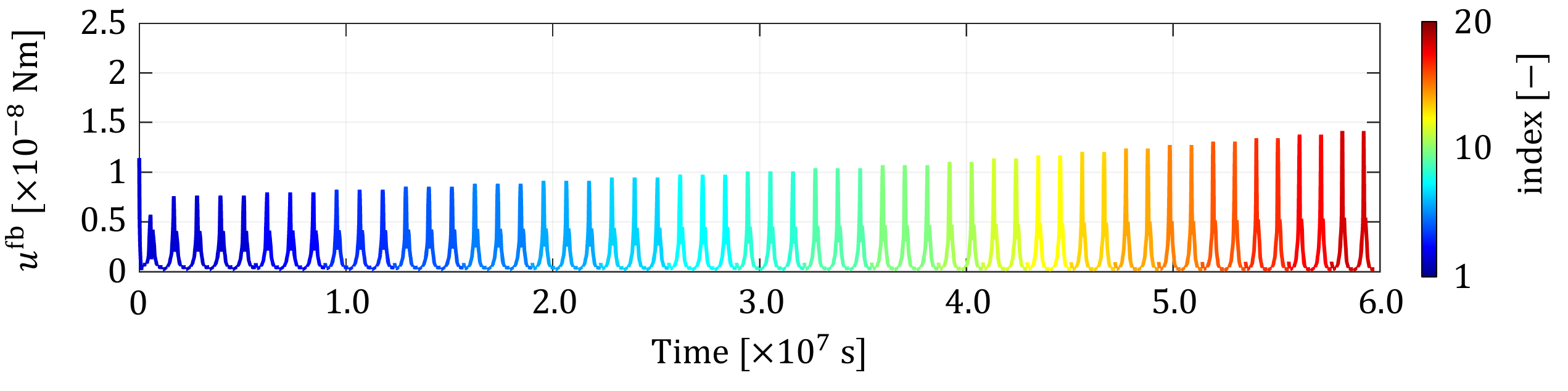}
		\caption{PD control}
		\label{fig:ufb_transfb}
	\end{subfigure}\\
	\begin{subfigure}{1\linewidth}
	\centering
		\includegraphics[width=0.85\hsize]{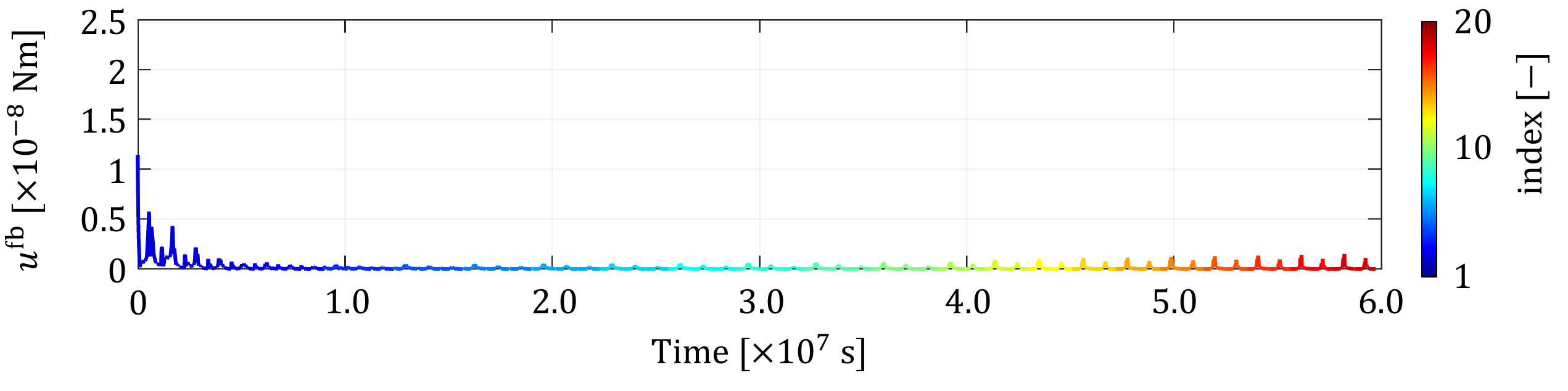}
     	\caption{PD control + RC}
	\end{subfigure}\\
 	\caption{Time history of feedback control input norm and disturbance torque norm.}
    \label{fig:ufbrc_transfb}\vspace{10pt}
    \begin{subfigure}{1\linewidth}
	\centering
		\includegraphics[width=0.85\hsize]{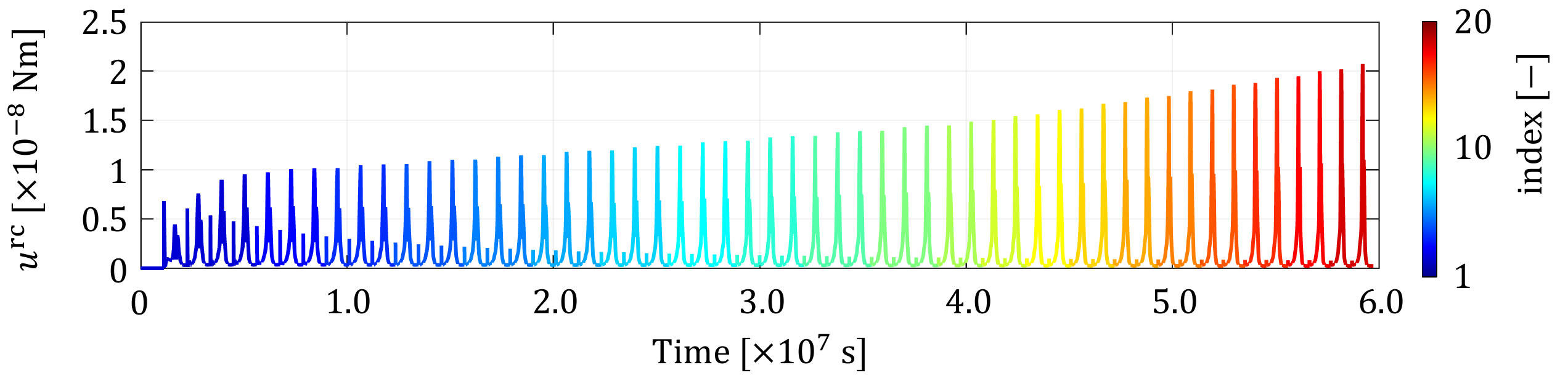}
  	\end{subfigure}\\
		\caption{Time history of RC input norm $\bm{u}^{\mathrm{rc}}$}
		\label{fig:urc_transfbrc}
\end{figure}

It is assumed that the only disturbance torque that affects the attitude motion of the spacecraft is the GG torque from the Earth and Moon, and that there is no precise orbital information available such as the GG torque. Figure \ref{fig:trans_GGT} shows the magnitude of the GG torque experienced by the spacecraft when it actually transfers orbits from index 1 to index 20. 
\begin{table}[htbp] 
\caption{Parameters of spacecraft}
   \centering
   \scalebox{1}{
   \renewcommand\arraystretch{1.5}
   \begin{tabular}{cccl}\hline\hline
     Parameters of spacecraft & Symbols & Values& \\ \hline 
     Inertia moment  & $J$ & $\mathrm{diag}(135,175,125)$ &[kgm$^2$] \\ 
     Uncertainty of Inertia moment & $\bar{J}$ & $\mathrm{diag}(121,157,112)$ &[kgm$^2$] \\  
     Initial angular velocity & $ \bm{\omega}_0$& $[0,0,0]^T$ &[-]  \\ 
     Initial quaternion values & $\bm{q}_0 $& $[0,0,0,1]^T$&[-]  \\ 
     Target angular velocity &$ \bm{\omega}_d$& $[0,0,0]^T$ &[-]  \\ 
     Target quaternion values & $\bm{q}_d $& $[-0.4,0.2,0.4,0.8]^T$&[-]  \\ \hline\hline
   \end{tabular}
   }
   \label{tb:parameter of spacecraft}
\end{table}
\begin{table}[htbp] 
\caption{Parameters of Numerical Analysis}
   \centering
   \scalebox{1}{
   \renewcommand\arraystretch{1.5}
   \begin{tabular}{cccl}\hline\hline
     Parameters & Symbols & Values& \\ \hline 
    Orbit transfer time (index:1$\sim$20) & $T^*$ & 1.59$\times 10^2$ &[-]  \\Orbit transfer time (index:1$\sim$20) & $T$ & 6.08$\times 10^7$ &[s]  \\ 
     Control cycle & $dt$ & $1.0\times 10^{-3}$& [-] \\ 
     Anticipatory interval & $\Delta$ & $2.0\times 10^{-2}$ &[-]  \\ 
     Number of trials (index:1) & $N$ & 5& [orbits] \\
     Number of trials (index:2$\sim$10) & $N$ & 3& [orbits] \\
     Number of trials (index:11$\sim$20) & $N$ & 2& [orbits] \\\hline\hline
   \end{tabular}
   }
   \label{tb:parameter of numerical analysis}
\end{table}
\begin{table}[htbp]
   \caption{Gain parameters}
   \centering
\renewcommand\arraystretch{1.5}
   \begin{tabular}{cc}\hline\hline
    Parameter &Values\\\hline
    $k^{\mathrm{fb}}_p$ &20 \\
    $k^{\mathrm{fb}}_d$&10  \\
    $k^{\mathrm{ILC}}_p$ &15 \\
    $k^{\mathrm{ILC}}_d$&8\\ \hline\hline
   \end{tabular}
   \label{tb:gain parameters}
\end{table}
Figure~\ref{fig:q4_transfb} shows the convergence results of the scalar part of the quaternion to the target value when the feedback input is used with fixed PD gains. Figure~\ref{fig:q4_transfbrc} shows the convergence results when a control input for attenuation of the GG torque by RC is added to the feedback input at fixed PD gains.
As shown in Fig.~\ref{fig:q4_transfb}, attitude control using only feedback input by the PD control [Eq.~\eqref{eq:Feedback}] cannot sufficiently attenuate the periodic disturbance torque, resulting in low convergence accuracy. To improve convergence accuracy, RC input is essential. 
Figure~\ref{fig:q4_transfbrc} shows the results of the combined PD control and RC [Eq.~\eqref{eq:input}] where RC input also acts as feedforward input. It can be seen that the GG torque remains attenuated even when the reference halo orbits change during the orbital transfer, indicating that the convergence accuracy is improved. Figure~\ref{fig:ufb_transfb} shows the time history of the feedback input by PD control and combined PD control and RC.
The feedback input alone is larger at proximity to the Moon to attenuate the effects of the GG torque. 
In the combined approach, the feedback input decreases as the orbit changes and the RC input increases to offset the effects of the GG torque. From Figs.~\ref{fig:ufbrc_transfb} and~\ref{fig:urc_transfbrc},  the RC input is determined from the effects of the disturbance torque experienced during the orbital motion one period earlier. Therefore, the RC input is increasing as the orbit period increases, as is the behavior of the GG torque. This implies that there is a direct correlation between the determination of RC input and the changes in the unknown periodic disturbance torque.
\subsection{Attitude control considering RWs dynamics}
\begin{figure}[htbp]
  \centering
    \includegraphics[scale=0.6]{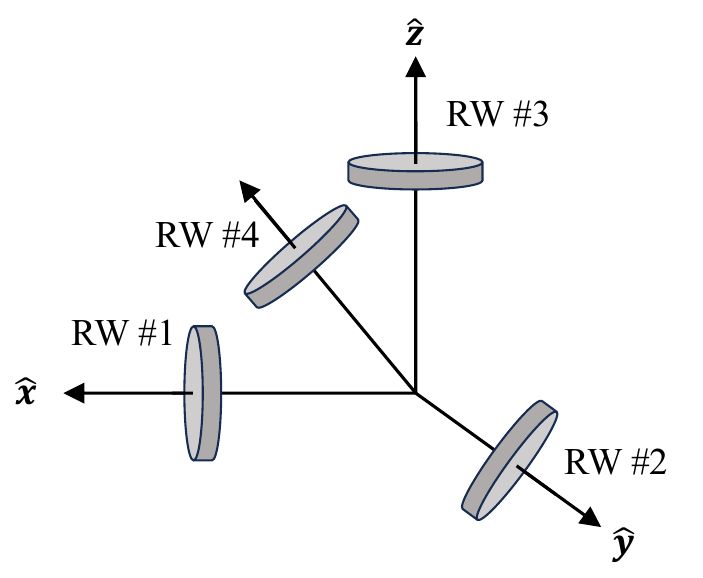}
	\caption{RW configuration}%
	\label{fig:RW Configuration}
\end{figure}

\begin{figure}[htbp]
  	\centering
		\includegraphics[scale =0.3]{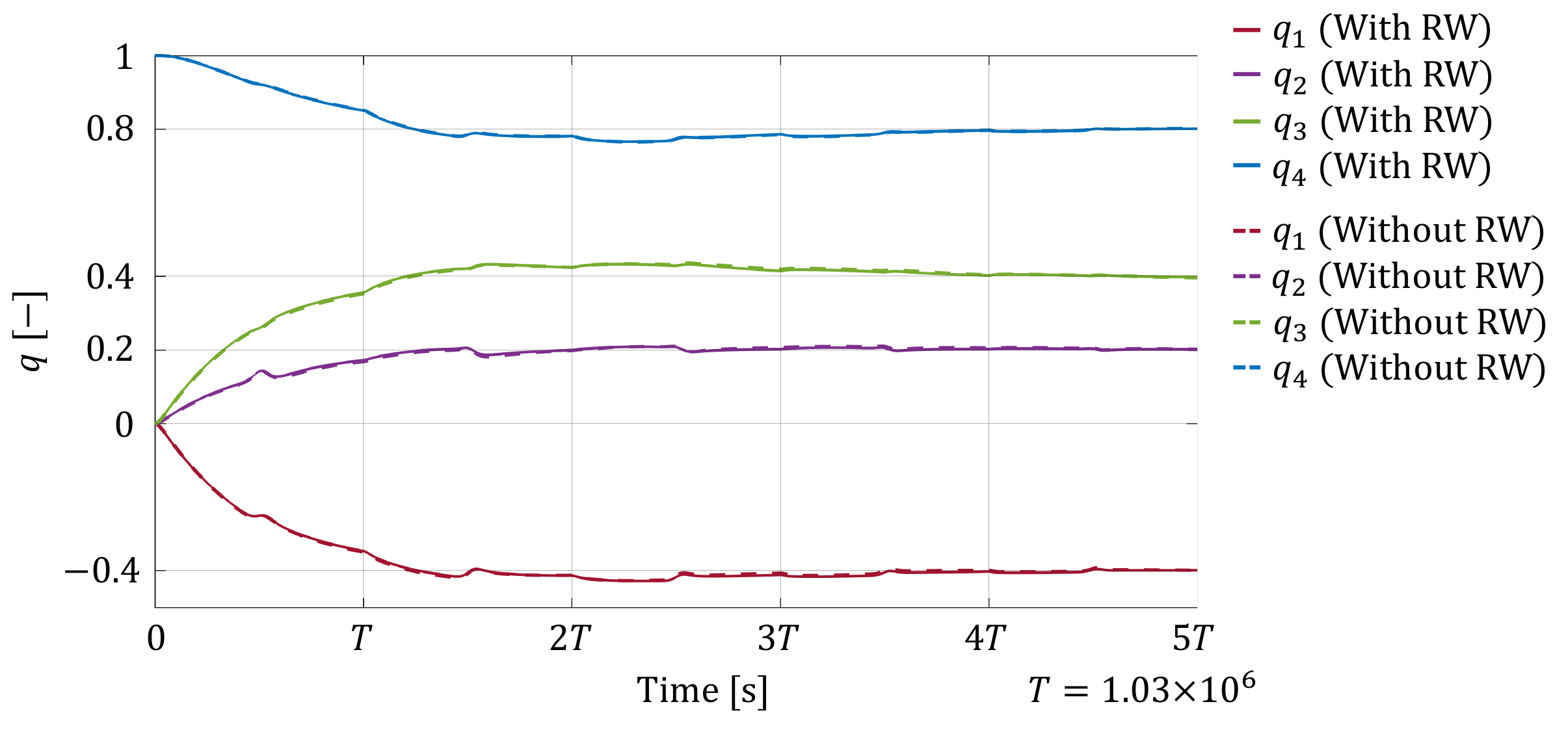}
		\caption{Time history of quaternion}
		\label{fig:uq_RW}
\end{figure}

\begin{figure}[htbp]
	\centering
	\begin{subfigure}{1\linewidth}
 	\centering
		\includegraphics[scale =0.3]{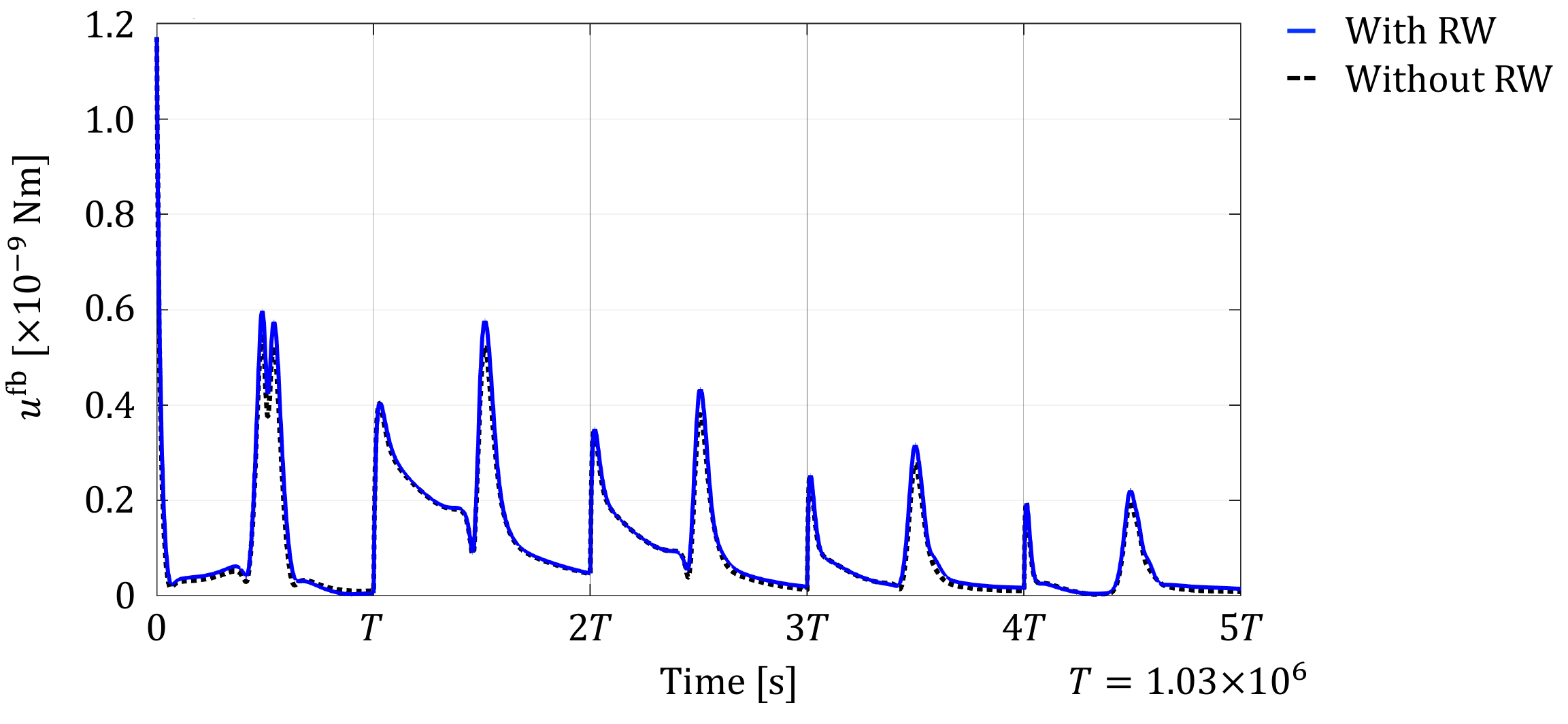}
		\caption{Time history of feedback inputs norm $\bm{u}^{\mathrm{fb}}$}
		\label{fig:ufb_RW}
	\end{subfigure}\\  \vspace{10mm}
 	\begin{subfigure}{1\linewidth}
  	\centering
		\includegraphics[scale =0.3]{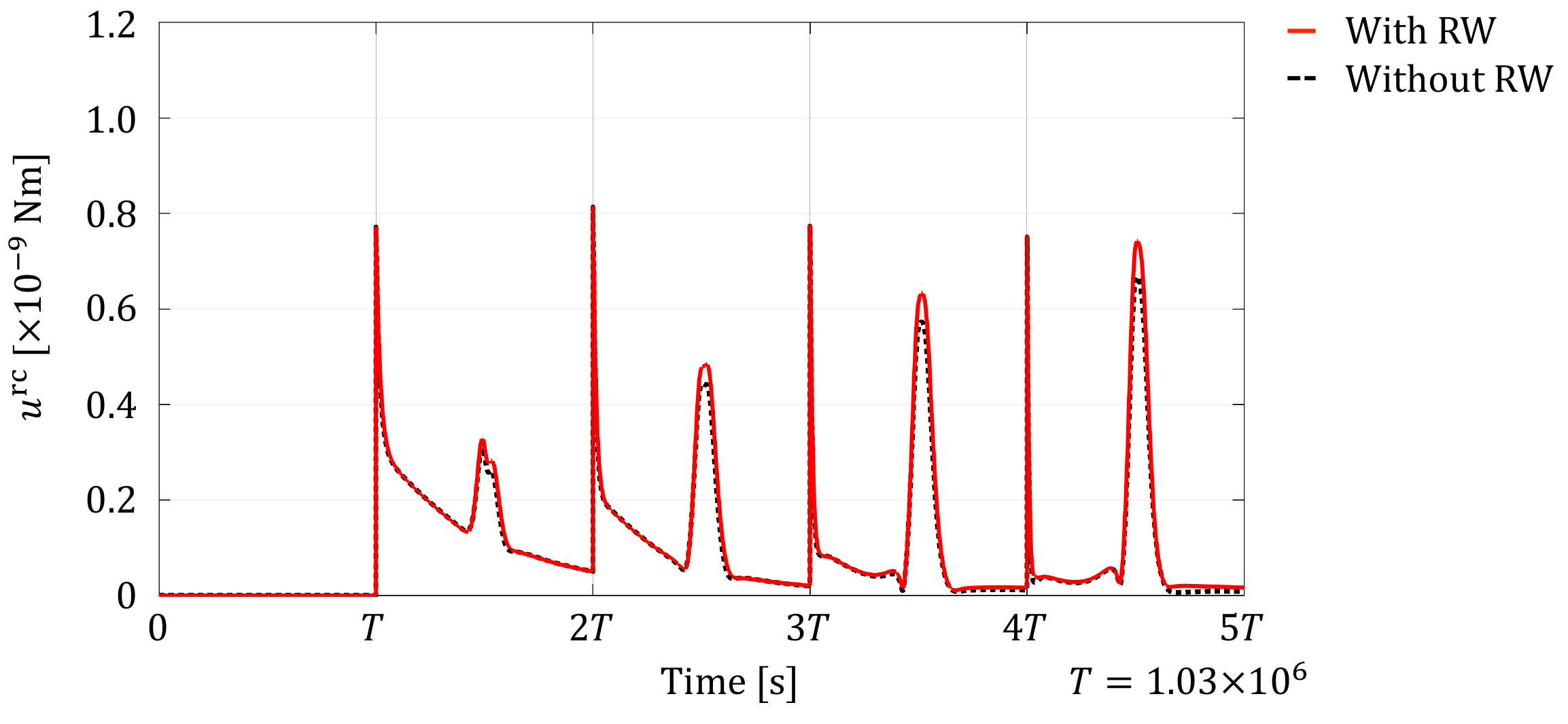}
		\caption{Time history of RC input norm $\bm{u}^{\mathrm{rc}}$}
		\label{fig:urc_RW}
	\end{subfigure}\\
	\caption{Time history of feedback inputs norm and RC input norm.}
	\label{fig:RW_fbrc_control}
\end{figure}

In this subsection, the coupled motion of the spacecraft moving on the reference orbit of index 20 in Fig.~\ref{fig:halofamily} is considered. Attitude control considering the RW dynamics, which is one of the internal force actuators. Regarding the RWs, the three-axis one-diagonal configuration of the four RWs~\cite{2015RW,Takahiro2016} shown in Fig.~\ref{fig:RW Configuration} is adopted. 
Given a target attitude angle, a torque command value is calculated from the spacecraft attitude angle and attitude angular velocity, and designed to follow the ideal control torque [Eq.~\eqref{eq:input}]. The total angular momentum of the spacecraft with RWs $\bm{H}_s$ is described as the sum of the angular momentum of the spacecraft and the four RWs as follows:
\begin{equation}
    \bm{H}_s=J\bm{\omega}+\bm{h}_{\mathrm{rw}}
    \label{eq:ang_Act}
\end{equation}
where $\bm{h}_{\mathrm{rw}}$ is the angular momentum of the four RWs given by
\begin{equation}
\bm{h}_{\mathrm{rw}}=G_sI_{\rm{rw}}\bm{\varOmega}
\end{equation}
where $I_{\rm{rw}}$ is the inertia moment of the wheel around the spin axis, $\bm{\varOmega}:=[\varOmega_1, \cdots, \varOmega_4]\in\bm{\mathbb{R}}^3$ is the wheel's angular velocity, and $G_s:=[\hat{\bm{s}}_1, \cdots, \hat{\bm{s}}_4]\in\bm{\mathbb{R}}^{3\times4}$ is the spin axis vector array of the four RWs. The parameters for the RW body and arrangement are shown in Table \ref{tb:RW}. In addition, information on the reference orbit and control cycle used in this numerical calculation are shown in Table \ref{tb:parameter of numerical analysis_RW}. The control gains for the PD control and RC are shown in Table \ref{tb:gain parameters_RW}.\\
The control torque of the RWs $\bm{u}^{\mathrm{rw}}$ is described as follows:
\begin{equation}
    \bm{u}^{\mathrm{rw}} = -G_sI_{\rm{rw}}\dot{\bm{\varOmega}}
    \label{eq:RW_Angvelo}
\end{equation}
The angular acceleration generated by RWs concerning the ideal control input torque required by the control law is described as follows:
\begin{equation}
    \dot{\bm{\varOmega}}=\frac{1}{I_{\rm{rw}}}{G}^T_s(G_s{G}^T_s)^{-1}\bm{\tau}_r
\end{equation}
where $\bm{\tau}_r$ is the torque command value given to RWs and is described by the combination of feedback input by PD control and RC input as follows:
\begin{equation}
    \bm{\tau}_r(t) =  \bm{u}^{\mathrm{fb}}(t)+\bm{u}^{\mathrm{rc}}(t)
    \label{eq:tau_r}
\end{equation}
\begin{table}[htbp]
   \caption{Parameters of RWs}
   \centering
   \scalebox{1}{
   \renewcommand\arraystretch{1.5}
   \begin{tabular}{cccl}\hline\hline
    Parameters of RW & Symbols & Values &\\ \hline
    Inertia moment (RW equipped)  & ${J}_{\rm{rw}}$ & diag$(10,10,8)$& [kgm$^2$] 
    \\Inertia moment  & $I_{\rm{rw}}$ & 0.002 &[kgm$^2$] \\  
     Dimensionless inertia moment & $I^*_{\rm{rw}}$ & $2.0\times10^{-4}$ &[-] \\ 
     Wheel's angular velocity & $\bm{\varOmega}_0$ & $100[1,1,1,-\sqrt{3}]^T$ &[rad/s] \\ 
     Spin axis vector$^1$ & $\hat{\bm{s}}_1$& $[1,0,0]^T$ &[-] \\
     Spin axis vector$^2$ & $\hat{\bm{s}}_2$& $[0,1,0]^T$ &[-] \\
     Spin axis vector$^3$ & $\hat{\bm{s}}_3$& $[0,0,1]^T$ &[-] \\
     Spin axis vector$^4$ & $\hat{\bm{s}}_4$& $1/\sqrt{3}[1,1,1]^T$ &[-] \\ \hline\hline
     \label{tb:RW}
   \end{tabular}
   }
\end{table}
\begin{table}[htbp] 
\caption{Parameters of Numerical Analysis}
   \centering
   \scalebox{1}{
   \renewcommand\arraystretch{1.5}
   \begin{tabular}{cccl}\hline\hline
     Parameters & Symbols & Values& \\ \hline 
    Orbit transfer time (index:1$\sim$20) & $T^*$ & 13.74$\times 10^2$ &[-]  \\
    Orbit transfer time (index:1$\sim$20) & $T$ & 5.16$\times 10^6$ &[s]  \\ 
     Control cycle & $dt$ & $1.0\times 10^{-3}$& [-] \\ 
     Anticipatory interval & $\Delta$ & $2.0\times 10^{-2}$ &[-]  \\ 
     Number of trials & $N$ & 5& [orbits]\\ \hline\hline
   \end{tabular}
   }
   \label{tb:parameter of numerical analysis_RW}
\end{table}
\begin{table}[htbp]
   \caption{Gain parameters}
   \centering
\renewcommand\arraystretch{1.5}
   \begin{tabular}{cc}\hline\hline
    Parameter &Values\\\hline
    $k^{\mathrm{fb}}_p$ &30 \\
    $k^{\mathrm{fb}}_d$&20  \\
    $k^{\mathrm{ILC}}_p$ &20 \\
    $k^{\mathrm{ILC}}_d$&8\\ \hline\hline
   \end{tabular}
   \label{tb:gain parameters_RW}
\end{table}
In the numerical simulation, the spacecraft repeats five orbits in the reference orbit. Figure~\ref{fig:uq_RW} shows the time history of the quaternion to the target values where the solid line is the result using the control torque of RW and the dashed line is the result using the ideal control inputs directly without RWs. 
Figure~\ref{fig:RW_fbrc_control} shows the time histories of the norm of the feedback and RC inputs with and without RW where the black dashed line represents the ideal control inputs without RW, while the blue and red lines represent the time history of the feedback and RC inputs with RW, respectively.
Considering the RW dynamics, there is a slight error in the initial trial, but it can be seen that the control torque of RW tracks the ideal control input, which is the target control torque, as the trial is updated. This small error is due to the time delay in transmitting the ideal control input to the RW, but the RW dynamics can respond quickly to the ideal control input. The results also show that the combined control method of PD control and RC can be directly applied to the attitude control law with or without the attitude control actuators, since the combined control method can control the attitude corresponding to the total angular momentum of the spacecraft $\bm{H}s$, which varies according to the angular momentum of the RW $\bm{h}{\mathrm{rw}}$ [Eq.~\eqref{eq:ang_Act}].

\begin{figure}[htbp]
	\centering
		\includegraphics[scale =0.3]{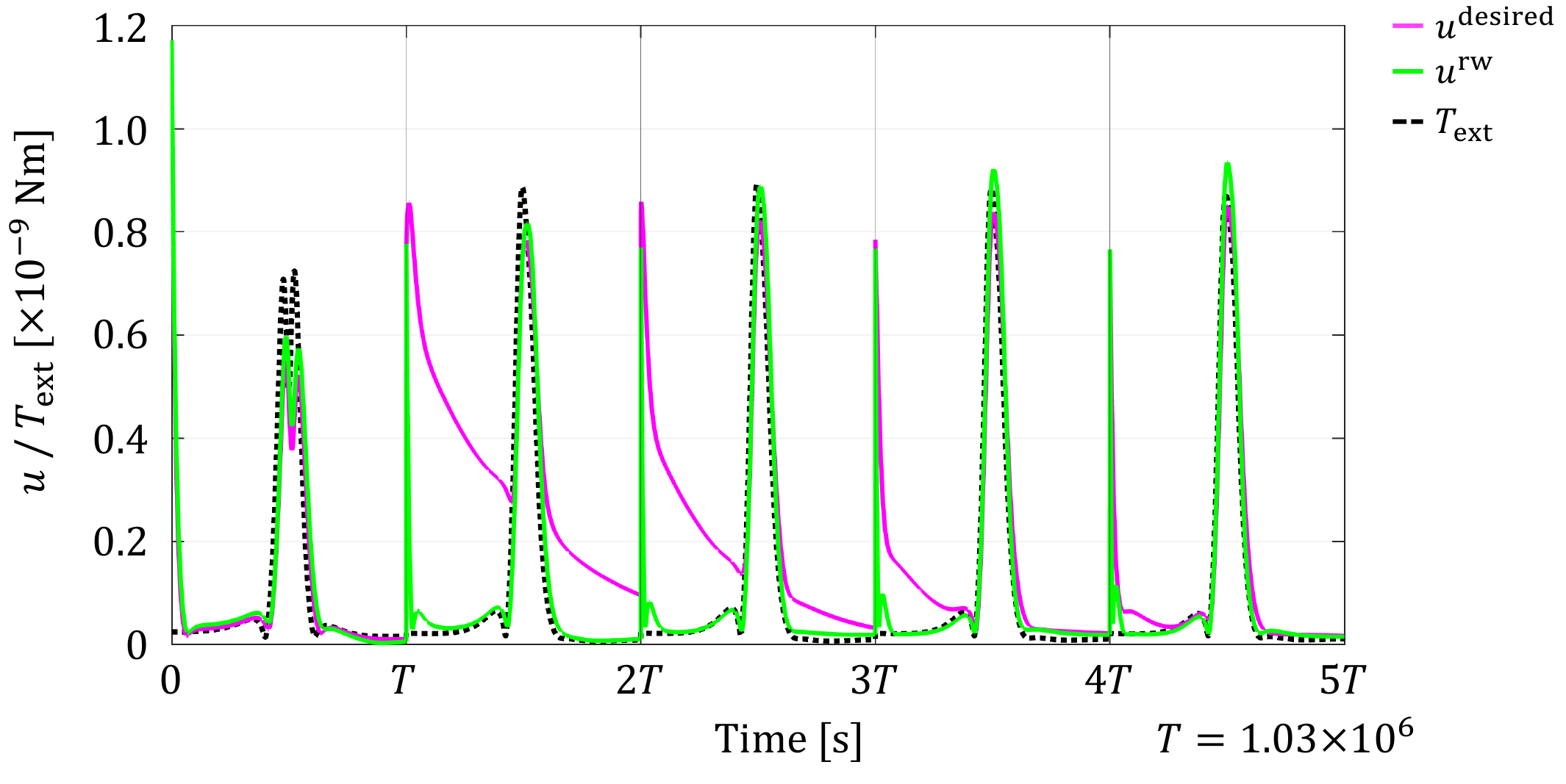}
		\caption{Time histories of control inputs norm and GG torque norm}
		\label{fig:ufb2_RW}
\end{figure}

Figure~\ref{fig:ufb2_RW} shows the time history of the GG torque norm where the black dashed line is the GG torque norm, the magenta solid line is the ideal control input norm (without RW), and the green solid line is the control torque of RW norm.
In the first period, the RW control torque does not follow the behavior of the GG torque. This is because feedback control dominates in the initial phase (0 to 2T), while the RC input, which is based on previous trials, has not yet accurately captured the unknown GG torque.
It can be seen that the ideal control input tracks the GG torque and the RW control torque tracks the ideal control input by continuing to learn through trial updates. These results show that the combination of feedback control and RC not only stabilizes the system but also adaptively adjusts and generates control inputs by learning the periodicity and dynamic characteristics of the periodic disturbance torque.\par
\begin{figure}[htbp]
  	\centering
		\includegraphics[scale =0.3]{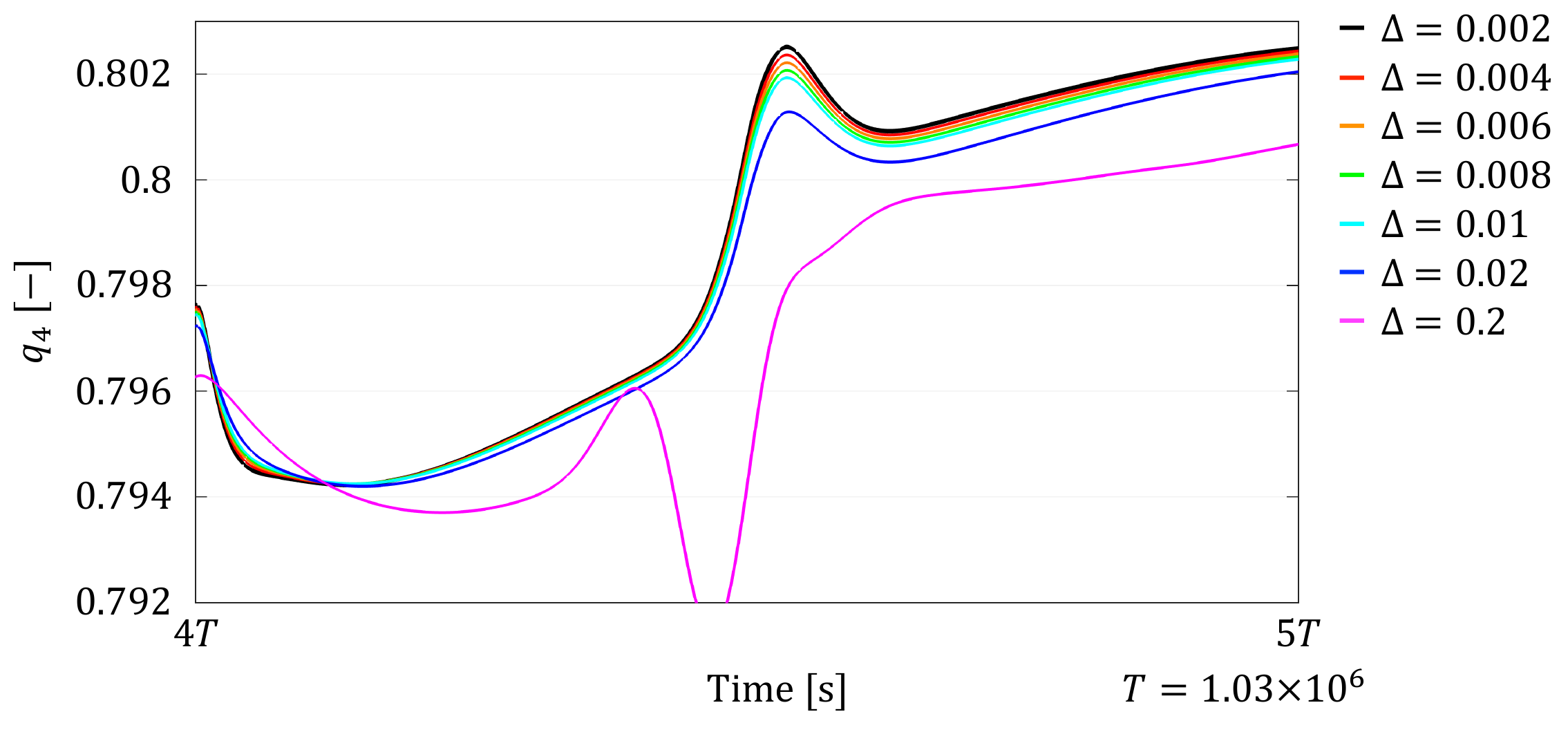}
		\caption{Time history of $q_4$}
		\label{fig:q4_delta}
\end{figure}

Next, the effect of the anticipatory parameter $\Delta$ [Eq.~\eqref{eq:RC scheme}] is examined. It is clear that the state quantity becomes unstable as the number of trials is increased because the control input of the RC diverges when the control gain of the RC is fixed and the delta is chosen such that the convergence condition [Eq.~\eqref{eq:theorem_rc}] is not satisfied.
Figure~\ref{fig:q4_delta} shows the time history of the scalar part of quaternion $q_4$ in the final trial (5T) when the delta is increased. The convergence speed is improved because the response to errors is faster as the delta is larger. However, it can be seen that choosing a $\Delta$ that does not satisfy the condition causes the attitude angle to be disturbed. This is because if $\Delta$ is large, the boundedness of the state is not guaranteed, and it diverges as the trials are repeated. Therefore, it is necessary to select an appropriate $\Delta$.
\par
\begin{figure}[htbp]
  	\centering
		\includegraphics[scale =0.3]{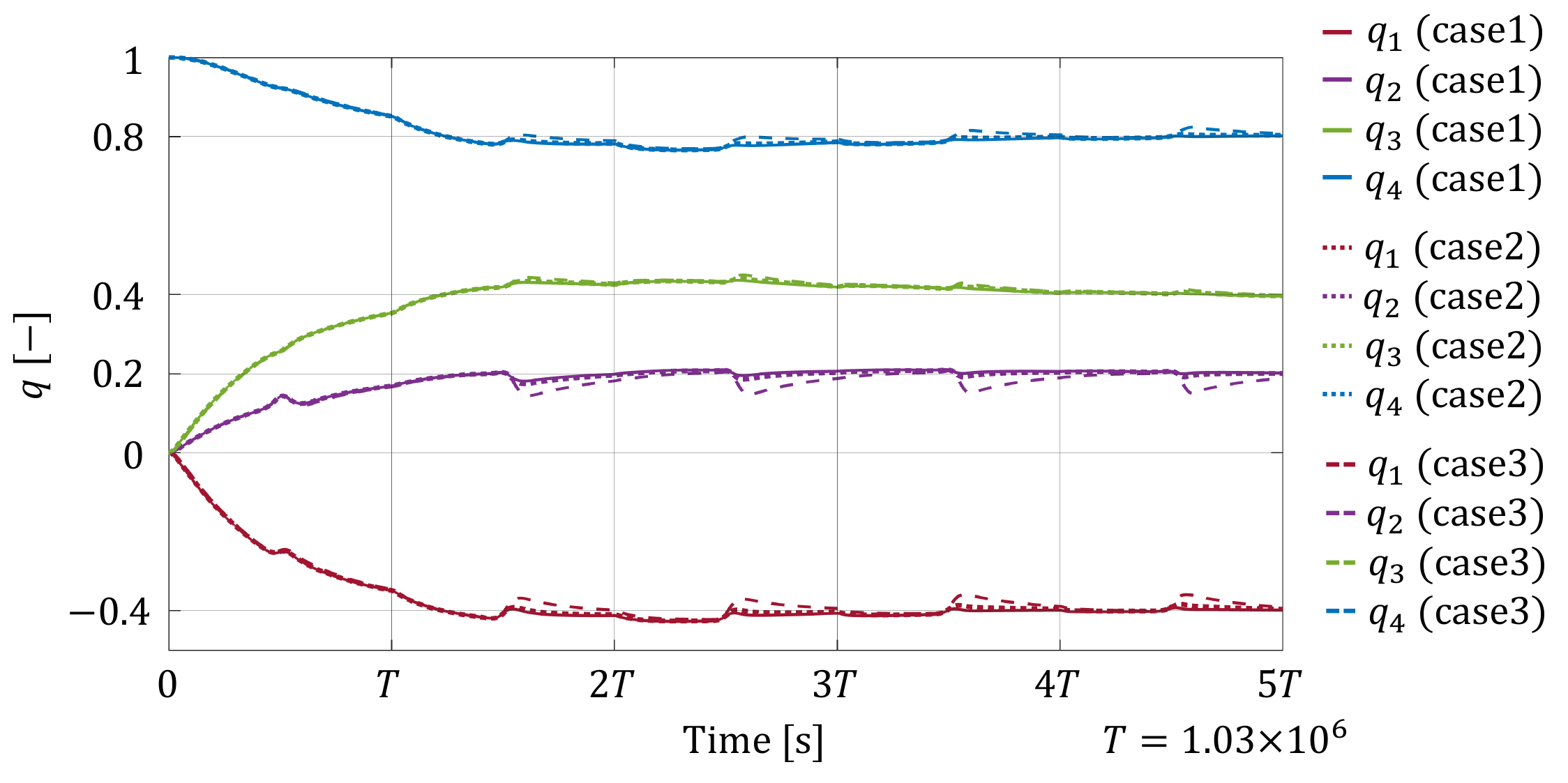}
		\caption{Time history of quaternion}
		\label{fig:q_case}
\vspace{10mm}
  	\centering
		\includegraphics[scale =0.3]{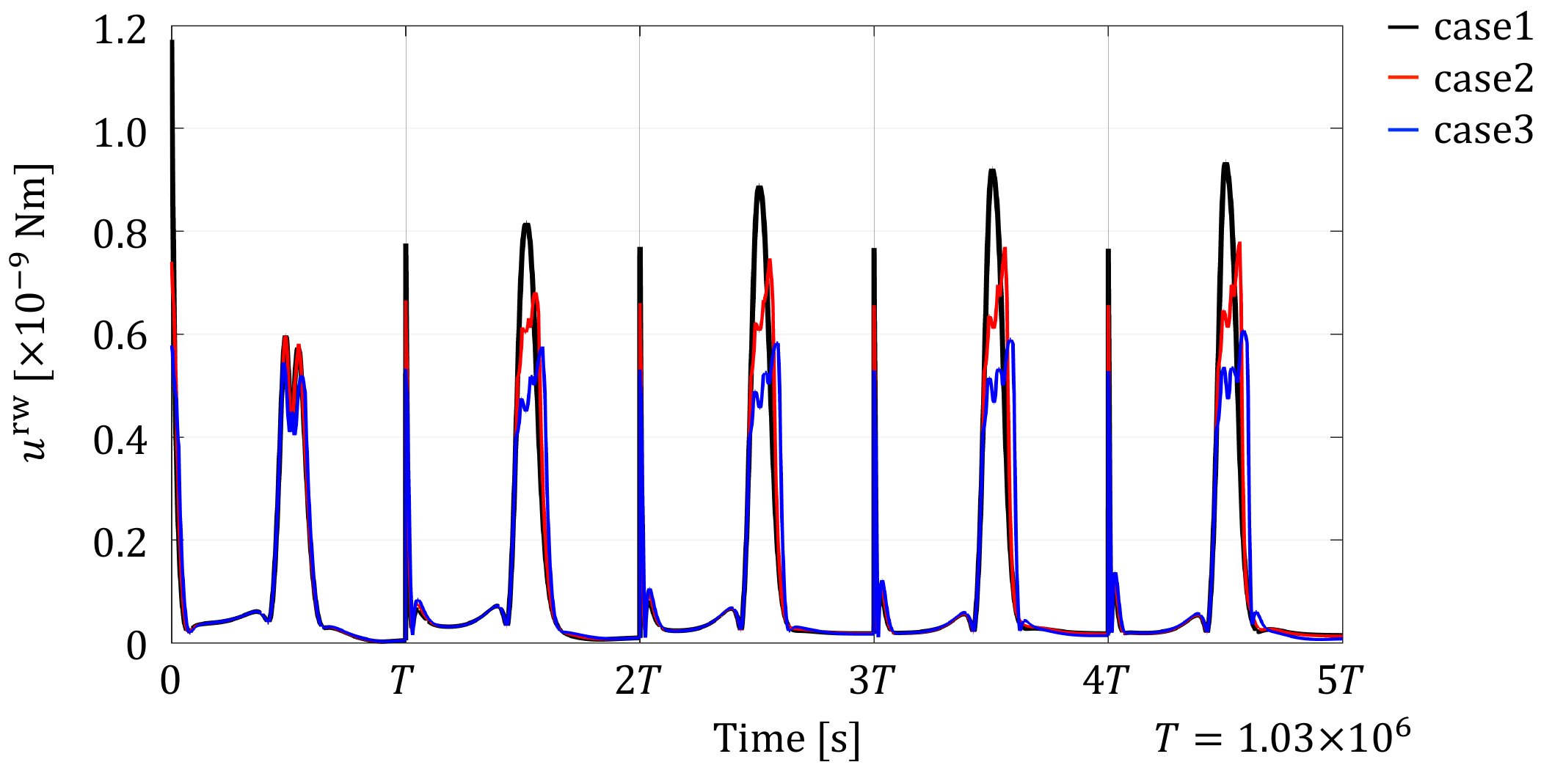}
		\caption{Time history of control torque of RWs}
		\label{fig:uRW_case}
\end{figure}

Finally, the spacecraft attitude and RC behavior are examined when the RW input is saturated. A constraint is set on the RW wheel angular velocity [Eq.~\eqref{eq:RW_Angvelo}]. 
Table ~\ref{tb:limit_input} shows the magnitude of the RW control torque with constraints in case 1 to 3. Figure ~\ref{fig:uRW_case} shows the time history of the RW control torque for each case. 
\begin{table}[htbp]
   \caption{Upper limit of the magnitude of control torque of RW}
   \centering
\renewcommand\arraystretch{1.5}
   \begin{tabular}{cc}\hline\hline
    case  & $u_{max}$\\ \hline
    case 1 & $\infty$ \\
    case 2 & $0.8\times10^{-9}~~[\mathrm{Nm}]$\\
    case 3 & $0.6\times10^{-9}~~[\mathrm{Nm}]$\\ \hline\hline
   \end{tabular}
   \label{tb:limit_input}
\end{table}
Figure ~\ref{fig:q_case} shows the time history of the quaternion with and without torque limitation for five orbits in the reference orbit for each case. Throughout the simulation, it can be seen that the RW control torque is generated to satisfy the magnitude constraint. The output torque is reduced due to the saturation constraint of the RC input, and the quaternion convergence is decreased by the inability to accurately attenuate the GG torque.
However, the quaternion converges to the target value as learning proceeds, except for small deviations during close Moon approaches. This is because input saturation does not change the stability condition of RC [Eq.~\eqref{eq:theorem_rc}]. 
In fact, assume $u^*(t) \in [-u_{max}, u_{max}]$ denotes the time history of ideal control input which satisfies
\begin{equation}
    ||\bm{u}^*(t) - \bm{u}(t)|| \leq ||u^* - \bm{u}(t-T)||
\end{equation}
Then, it follows that
\begin{equation}
    ||\bm{u}^*(t) - \bm{u}(t)|| \leq ||u^* - \bm{u}^{sat}(t-T)||\leq ||u^* - \bm{u}(t-T)||
\end{equation}
from the relation
\begin{equation}
   || \bm{u}^{sat}(t-T)||= ||\max\{ \bm{u}(t-T), u_{max} \}|| \leq  ||\bm{u}(t-T)||
\end{equation}
where $\bm{u}^{sat}$ is the control input under input saturation of RWs.
Therefore, RC with input saturation can be discussed in a similar manner as the RC without input saturation. 
The simulation results demonstrate the ability of RC  that can improve the convergence accuracy of the quaternion step by step until saturation is achieved by repeating the trials in the case with input saturation.
\section{Conclusion}
This paper presented an anticipatory RC framework for the attitude control of a spacecraft along a halo orbit near the Earth–Moon L$_2$ point. The proposed method achieves robust and adaptive control performance under periodic disturbances, particularly GG torques that are difficult to model precisely. By exploiting the periodicity of the system dynamics and incorporating future error prediction, the anticipatory RC enables accurate disturbance attenuation and faster convergence to the target attitude. Numerical simulations validated the method’s effectiveness, showing that the spacecraft maintains stable attitude control even during orbit transfers with dynamically changing reference trajectories. Furthermore, the proposed control law generates accurate torques within Reaction Wheel limits. These results confirm the RC-based control framework as a practical approach for spacecraft attitude control, capable of addressing the challenges posed by periodic disturbances and actuator dynamics in multi-body environments.
\bibliography{main}
\end{document}